%% file: main.tex
\documentclass[sigconf,screen]{acmart}
\usepackage[utf8]{inputenc}
\usepackage{xcolor,colortbl}
\usepackage{hhline}

\acmConference[SESAME'24]
{The 2nd Workshop on Serverless Systems, Applications and Methodologies}
{22-25 April 2024, Athens, Greece}

\begin{abstract}
\input{abstract}
\end{abstract}

\title{Imaginary Machines: A Serverless Model for Cloud Applications}

\author{
Michael Wawrzoniak$^1$,
Rodrigo Bruno$^2$,
Ana Klimovic$^1$,
Gustavo Alonso$^1$ \\
$^1$Systems Group, Dept. of Computer Science, ETH Zurich  \\
$^2$INESC-ID/Técnico, U. Lisboa
}

\begin{document}
\maketitle
\input{introduction}

\input{model}

\input{conclusion}

\bibliographystyle{ACM-Reference-Format}
\bibliography{paper}

\end{document}

%% file: abstract.tex
Serverless Function-as-a-Service (FaaS) platforms provide applications with resources that are highly elastic, quick to instantiate, accounted at fine granularity, and without the need for explicit run-time resource orchestration.
This combination of the core properties underpins the success and popularity of the serverless FaaS paradigm.
However, these benefits are not available to most cloud applications because they are designed for networked virtual machines/containers environments.
Since such cloud applications cannot take advantage of the highly elastic resources of serverless and require run-time orchestration systems to operate, they suffer from lower resource utilization, additional management complexity, and costs relative to their FaaS serverless counterparts.

We propose \emph{Imaginary Machines}, a new serverless model for cloud applications. 
This model (1.) exposes the highly elastic resources of serverless platforms as the traditional network-of-hosts model that cloud applications expect, and (2.) it eliminates the need for explicit run-time orchestration by transparently managing application resources based on signals generated during cloud application executions. %
With the Imaginary Machines model, \emph{unmodified cloud applications become serverless applications}.
While still based on the network-of-host model, they benefit from the highly elastic resources and do not require runtime orchestration, just like their specialized serverless FaaS counterparts, promising increased resource utilization while reducing management costs.

%% file: introduction.tex
\section{Introduction} %

Cloud users today can choose between two main options for developing and managing their applications: renting VMs on-demand or using providers' serverless compute and storage services.
VMs come with a familiar programming/execution environment of network-of-hosts, but provide a lower level of resource elasticity, and users need to explicitly manage the infrastructure at run-time (e.g., using cluster managers, turn machines on/off based on the application load.)
Serverless platforms automate infrastructure management and provide highly elastic resources but use a restrictive FaaS programming model with limitations that reduce its broader applicability~\cite{HellersteinCIDR19}. To overcome the limitations, researchers have been exploring new serverless platform designs~\cite{Cloudburst}, additional infrastructure~\cite{pocket}, and constructing new FaaS-specialized applications~\cite{Lambada}.

This paper explores how to get the best of both worlds -- applying the serverless paradigm to cloud applications designed for networked VMs with the familiar network-of-hosts programming model. We explore how to automate infrastructure orchestration à la serverless for cloud applications.
We propose a serverless model for cloud applications, which we call \textit{Imaginary Machines}, that preserves the familiar, well-loved network-of-hosts programming model of VM-based applications with the orchestration automation of serverless platforms. We also discuss approaches to realize this model and propose an overlay approach based on an evolution of Boxer~\cite{Boxer-CIDR21} to realize the model on publicly available FaaS resources.

\section{Serverless Cloud Applications}
We believe that cloud applications based on the network-of-hosts model can also be serverless, taking advantage of highly elastic resources and automated infrastructure orchestration. Our perspective is based on the following observations and insights.
\\

\noindent
{\bf Serverless resources can suit cloud applications: } Although commonly viewed as lightweight and restricted, resources made available by FaaS can match many cloud application requirements. FaaS platforms continue to reduce limitations and increase resource limits, but even today's public platforms, such as AWS Lambda~\cite{awslambda}, can match many cloud application requirements of memory, compute capacity, state persistence, and reliability for some use cases.

Resources available in today's FaaS platforms have been increasing in size to the point where functions available today are comparable to the largest VMs available 15 years ago (memory and cpu), the environment in which many of the popular cloud applications have been originally designed for~\cite{zookeeper-start-date, spark-start-date}. AWS Lambda functions can be configured up to 10GB of memory and 6vCPU cores~\cite{aws-lambda-10GB-ram-announce}, matching largest AWS EC2 CPU instances introduced 12 years earlier~\cite{aws-high-cpu-announce}.

Second, the wide adoption of disaggregated storage resulted in cloud applications that are structured into multiple tiers of functionality, where many services do not depend on the local persistent state, which makes these layers a match for the ephemeral state provided by publicly available FaaS functions. For example, a completely unmodified application logic tier of a popular DeathStarBench~\cite{DeathStarBench} microservice benchmark runs in AWS Lambda~\cite{boxer-rapid}.

Third, FaaS functions are becoming more reliable and are no longer short-lived. The maximum execution time of FaaS functions has been increasing, in AWS Lambda, it is now 15 minutes~\cite{aws-lambda-15min-announce}.
\\

\noindent
{\bf Unbundling serverless resources from programming model: }
Serverless resources do not have to be bundled with the FaaS programming model, even in mainstream publicly available serverless platforms such as AWS Lambda. An evolution of Boxer~\cite{Boxer-CIDR21}, a platform providing the network-of-hosts programming model on top of AWS Lambda FaaS, has been demonstrated to run unmodified off-the-shelf distributed data processing systems such as Apache Spark and Drill~\cite{Boxer-CIDR24}, unmodified Apache Zookeeper, and the logic layer of the DeathStarBench microservice benchmark~\cite{shortlived-datacenters, boxer-rapid}.
These unmodified cloud applications run on AWS Lambda functions just as if they were running on VM-based networked hosts, each function instance having a routable network address, a resolvable hostname, and function-to-function network transport.

These projects demonstrated the feasibility of using the network-of-hosts model with serverless resources. However, they also took a step back by reintroducing the need for a serverful orchestration system (docker-compose~\cite{docker-compose} was used in~\cite{Boxer-CIDR24}) to manage executions of individual functions.
\\

\noindent
{\bf Serverless orchestration for cloud applications: }
We believe that automated infrastructure orchestration, an essential property of serverless, can also be realized for network-of-hosts model cloud applications.
We first describe a perspective on automated resource orchestration of existing FaaS and then propose how to achieve an analogous mechanism for the network-of-hosts model. 

The FaaS programming model enables automated infrastructure orchestration by embedding resource allocation signals in application control flow. The two signals generated during application execution are 
(1.) function invocation resulting in resource allocation 
of a function instance (and invocation of the function event handler), and 
(2.) function handler exit, when resources are released. 
As a ubiquitous and essential optimization, the resources are normally not released immediately after function handlers exit, instead, to provide low-latency warm-starts, function instances are first suspended for a limited time before they are terminated. 
FaaS applications are factored into collections of functions (possibly structured as dataflows~\cite{aws-step-functions}) that compose by invoking each other (directly or by producing events) based on internal control flow. Each such function invocation is a resource allocation signal used by FaaS platforms to dynamically rescale resources and create the illusion of avoiding resource orchestration.

Cloud applications based on the network-of-hosts model execute processes on hosts that 'compose' by communicating with each other via a network.
The analogous resource allocation signals to those embedded in FaaS functions (above) are also present in cloud applications.
Generally, any signal that indicates that any application process on a host function could be runnable is potentially %
a resource allocation signal for that host function (analogous to a function invocation).
Similarly, any signal that indicates that all application processes of a host are idle is a resource release signal for that host function (analogous to a function event handler exiting).
To achieve the automated infrastructure orchestration for the network-of-hosts model, the platform must observe such signals and perform the necessary on-demand resource allocation actions of allocating and releasing host functions, just like in FaaS.

Although not as explicit as in FaaS, cloud applications provide such signals; these signals are already used by local operating systems of running hosts to perform local application process scheduling (e.g., in response to a signal indicating that data is available for a waiting blocked process, the scheduler may choose to make the process runnable).
However, in contrast to VM-based infrastructure, to achieve serverless fine-grain scaling, host functions of cloud applications must be allocated on demand only when needed and released (suspended or terminated) when not used.
This means that, instead of the host's local operating system, which may be suspended or not even started, the appropriate resource allocation signals must be observed and acted on \emph{externally} to the host function.
Fortunately, assuming that cloud applications run entirely on the serverless platform, the allocation signals for all non-running host functions must be generated by (and observable at) the running host functions or external network requests, all observable by the serverless platform.
As a concrete example, when application processes initiate network communication to a valid destination host that is not running (e.g., during initialization or scaling up), the platform must use this signal to allocate the new destination host function automatically and in a transparent way to the already running application processes. 
Conversely, as application processes on a running host become idle, the platform suspends and eventually possibly terminates the host, automatically releasing unused application resources.
More signaling scenarios must be handled, which we do not describe here, but based on this approach, the resources used by network-of-hosts applications are allocated and released on-demand and transparently to the application, providing a form of automated infrastructure orchestration.

In order to provide application transparency, the resource allocation latencies (time to start or resume a host) must be sufficiently short (and infrequent) to be within tolerable network latencies of the applications.
Fortunately, as evidenced by publicly available systems such as AWS Lambda, latency to instantiate new host functions (cold $\sim$200ms) can be in the range of wide-area network latencies, suggesting they are within the tolerable range of connection timeouts that many cloud applications can handle.

%% file: model.tex
\section{Imaginary Machines Model}
We refer to the serverless execution model of cloud applications as the Imaginary Machines (IM) model.
\begin{itemize}
    \item \emph{From the application perspective}, the Imaginary Machines model presents a network-of-hosts programming model, where all possible (configured) hosts are already instantiated.
    Cloud application processes run \emph{imagining} as if all network destinations were already running and available.
    
    Significantly, the model \emph{does not} include any explicit run-time infrastructure orchestration just like in the FaaS model). 

    \item \emph{From the platform perspective}, the system must support the programming model that the applications expect (above) %
    while performing on-demand, automatic, transparent, and fine-grained resource orchestration for the application.
\end{itemize}

We are in the process of realizing a version of the IM model by extending Boxer~\cite{boxer-rapid} serverless overlay system that already provides network-of-hosts model on top of AWS Lambda FaaS by also including the necessary function allocation mechanism.

%% file: conclusion.tex
\section{Conclusion}
We motivated and described an alternative serverless model for cloud applications called Imaginary Machines.
Given that the dominant programming model in the cloud is based on the network-of-hosts model, providing applications based on that model with access to serverless resource elasticity and reduced operational complexity has the potential for a great impact.